\documentclass[conference]{IEEEtran}
\IEEEoverridecommandlockouts
\usepackage{cite}
\usepackage{graphicx}
\graphicspath{{./}{./pic/}}
\DeclareGraphicsExtensions{.pdf,.jpeg,.png}
\usepackage{float}
\usepackage{amsmath,amsfonts}

\usepackage{enumerate}
\usepackage{multirow}
\usepackage{url}
\usepackage{hyperref}
\usepackage{color}
\usepackage{booktabs}
\usepackage{threeparttable}
\usepackage{bm}

\hypersetup{linkcolor=green}
\newcommand{\RNum}[1]{\uppercase\expandafter{\romannumeral #1 \relax}}

\begin{document}
\title{Learning Environmental Sounds with Multi-scale \\ Convolutional Neural Network}

\author{\IEEEauthorblockN{Boqing Zhu$^1$, Changjian Wang$^{1}$, Feng Liu$^1$, Jin Lei$^1$, Zengquan Lu$^2$, Yuxing Peng$^{1}$}
\IEEEauthorblockA{$^1$\textit{Science and Technology on Parallel and Distributed Laboratory, National University of Defense Technology}\\
$^2$\textit{College of Meteorology and Oceanology, National University of Defense Technology}\\
Changsha, China \\
zhuboqing09@nudt.edu.cn}
}
% \author{\IEEEauthorblockN{Boqing Zhu, Changjian Wang, Jin Lei, Feng Liu, Yuxing Peng\thanks{Corresponding author.}}
% \IEEEauthorblockA{\textit{Science and Technology on Parallel and Distributed Laboratory, National University of Defense Technology}\\
% Changsha, China \\
% \{zhuboqing09\}@nudt.edu.cn; c\_j\_wang@yeah.net; \{leijin16, richardlf\}@nudt.edu.cn; pengyuxing@aliyun.com}
% }
\maketitle

\begin{abstract}
Deep learning has dramatically improved the performance of sounds recognition. However, learning acoustic models directly from the raw waveform is still challenging. Current waveform-based models generally use time-domain convolutional layers to extract features. The features extracted by single size filters are insufficient for building discriminative representation of audios. In this paper, we propose multi-scale convolution operation, which can get better audio representation by improving the frequency resolution and learning filters cross all frequency area. For leveraging the waveform-based features and spectrogram-based features in a single model, we introduce two-phase method to fuse the different features. Finally, we propose a novel end-to-end network called WaveMsNet based on the multi-scale convolution operation and two-phase method. On the environmental sounds classification datasets ESC-10 and ESC-50, the classification accuracies of our WaveMsNet achieve 93.75\% and 79.10\% respectively, which improve significantly from the previous methods.
% \vspace{3mm}
\end{abstract}

\section{Introduction} % (fold)
\label{sec:introduction}

The environmental sounds are a wide range of everyday audio events. The problem of environmental sound classification (ESC) is crucial for machines to understand surroundings. It is a growing research \cite{ESC_Dataset,Piczak2015Environmental,Dai2016Very,Salamon2014A,Tokozume2017Learning,Boddapati2017Classifying,Zhang2017Dilated} in the multimedia applications.

Deep learning has been successfully applied to this task and has generally achieved better results than traditional methods such as random forest ensemble \cite{randomForest} or support vector machine (SVM) \cite{ESC_Dataset}. The majority of models \cite{Piczak2015Environmental,Tokozume2017Learning} use spectrogram representation as input, such as log-mel features which compress the amplitude with a log scale on mel-spectrograms. The representations are often transformed further into more compact forms of audio features (e.g. MFCC \cite{Boddapati2017Classifying,Salamon2014A}) depending on the task. All of these processes are designed based on acoustic knowledge or engineering efforts and features might not suit with the classifier well.

Recently increasing focus has extended the end-to-end learning approach down to the level of the raw waveform and features could be learned directly from raw waveform rather than designed by experts. A commonly used approach is using a time-domain convolution operation on the raw waveform to extract features as audio representation for classification. Many of them \cite{Dai2016Very,Tara2015cldnn,Tara2016vad} matched or even surpassed the performances which employ spectral-based features. Papers \cite{Tokozume2017Learning,Tara2015front-end} have found the complementarity between waveform features and analytic signal transformation (e.g. log-mel). By combining these two kinds of features, they got a notable improvement on classification accuracy. These works just trained several independent models and calculated the average of output probabilities of each model.

However, there are still two deficiencies in the existing methods. Firstly, in the previous methods, fixed size filters were employed on the time-series waveform to extract features. However, it is always a trade-off for choosing the filter size. Wide windows give good frequency resolution, but does not have sufficient filters in the high frequency range. Narrow windows can learn more dispersed bands but get low frequency resolution \cite{Dai2016Very}. The feature extracted by single size filters might be insufficient for building discriminative representation under this dilemma. Secondly, the average method can not make full use of the complementary information in the waveform features and spectrogram features. It remains to be seen whether it can learn a combination automatically of different features in a single mode.

To address these two issues, we propose a novel multi-scale convolutional neural network (WaveMsNet), that can extract features by filter banks at multiple different scales and then fuse with the log-mel features in the same model. We show that our multi-scale CNN outperforms the single-scale models around 3\% on classification accuracy with same filters number, which is currently the best-performing method using merely waveform as input. After employing our proposed feature fusion method, the accuracy of classification is further improved. In summary, the unique contributions of this paper are threefold:

\vskip 0.1 cm
\begin{enumerate}[\indent 1.]
\setlength{\itemsep}{2pt}
\item  We analyze the inherent deficiencies of single-scale method on features extraction and propose a novel multi-scale time-domain convolution operation which can extract more discriminative features by improving the frequency resolution and learning filters cross all frequency area.
\item We propose a method of feature fusion which can combine spectral-based features with waveform features in a single model.
\item We design a new multi-scale convolutional neural network (WaveMsNet) based on above two, composed of filters with different sizes and strides. It outperforms the single-scale network and achieves the state-of-the-art model using only waveform as input. The classification accuracy has been further improved using our features fusion method.

\end{enumerate}
\vskip 0.1 cm

The remainder of this paper is organized as follows. Section 2 discusses related work. Section 3 gives our method and network’ architecture. Section 4 presents the experimental process and results, also, we analyze the results in this section. Finally, Section 5 concludes this paper.
% section introduction (end)

\section{Related Work}
\label{sec:related_work}
For years, designing an appropriate feature representation and building a suitable classifier for these features have used to be treated as separate problems in the sound classification task \cite{Tradition1,Tradition2,Tradition3,Tradition4,Tradition5,Tradition6}. For example, acoustic researchers have been using zero-crossing rate and mel-frequency cepstral coefficients (MFCC) as features to train a random forest ensemble \cite{randomForest} or support vector machine (SVM) \cite{SVMforspeaker,ESC_Dataset}. In these methods, classification stage is separated with feature extraction so that the designed features might not be optimal for the classification task.

Recently, deep learning based classifiers \cite{Salamon2016Deep} are used for the ESC task \cite{ESC_Dataset}. In particular, Convolutional Neural Network (CNN) has been observed to work better for this problem\cite{Tokozume2017Learning,Salamon2016Deep}. Since CNN classifier is useful for capturing the energy modulations across time and frequency-axis of audio spectrograms, it is well suited as classifier for ESC task\cite{Salamon2016Deep}. The spectral-based features, such as MFCC\cite{Salamon2014A}, GTCC\cite{Vacher2014Sound}, and TEO-based GTCC\cite{Agrawal2017Novel} are commonly used as input to extract more abstract features in the ESC task. In \cite{Piczak2015Environmental}, Piczak proposed a CNN which once was state-of-the-art method of ESC task using log-mel features and deltas (the first order temporal derivative) as a 2-channel input. 

End-to-end learning approach has been successfully used in image classification \cite{AlexNet,VGG,ResNet,Huang2016DenseNet,Ronneberger2015UNet,zhang2017airport,wang2018local} and text domain \cite{Zhang2015Character,Kim2015Character}. In the audio domain, learning from raw audio has been explored mainly in the automatic speech recognition (ASR) task \cite{Tara2015cldnn,Tara2015front-end,Tara2016vad}. They reported that the performance can be similar to or even superior to that of the models using spectral-based features as input. End-to-end learning approach has also been applied to music auto-tagging tasks as well \cite{Lee2017Sample}.  In \cite{Tokozume2017Learning}, Tokozume employed an end-to-end system on the ESC task for the first time, but the approach used traditional fixed filter size and stride length. More abundant features would be learned. 

In the papers \cite{Tara2015front-end,Tokozume2017Learning}, learned features can be fixed with another features at train time. Authors found noticeable improvements by supplementing log-mel filter banks features. They pre-trained two individual models which used raw wave and log-mel as input respectively and calculate the prediction of each window for probability-voting using the average of the output of these two networks. We try to explore an efficient method to combine these two kinds of features in one model in this paper.

Further, SoundNet \cite{Aytar2016SoundNet} proposed to transfer knowledge from visual models for sound classification. They used CNN models trained from visual objects and scenes to teach a feature extractor network for audio. However, it used a large number of external data, include many video clips. We hope to find an effective way to learn directly from audio.
% section related_work (end)

\section{Methods}
\subsection{Multi-scale convolution operation}
\begin{figure}[tb]
\setlength{\abovecaptionskip}{0pt}
\setlength{\belowcaptionskip}{10pt}
\centering
\includegraphics[scale=.5]{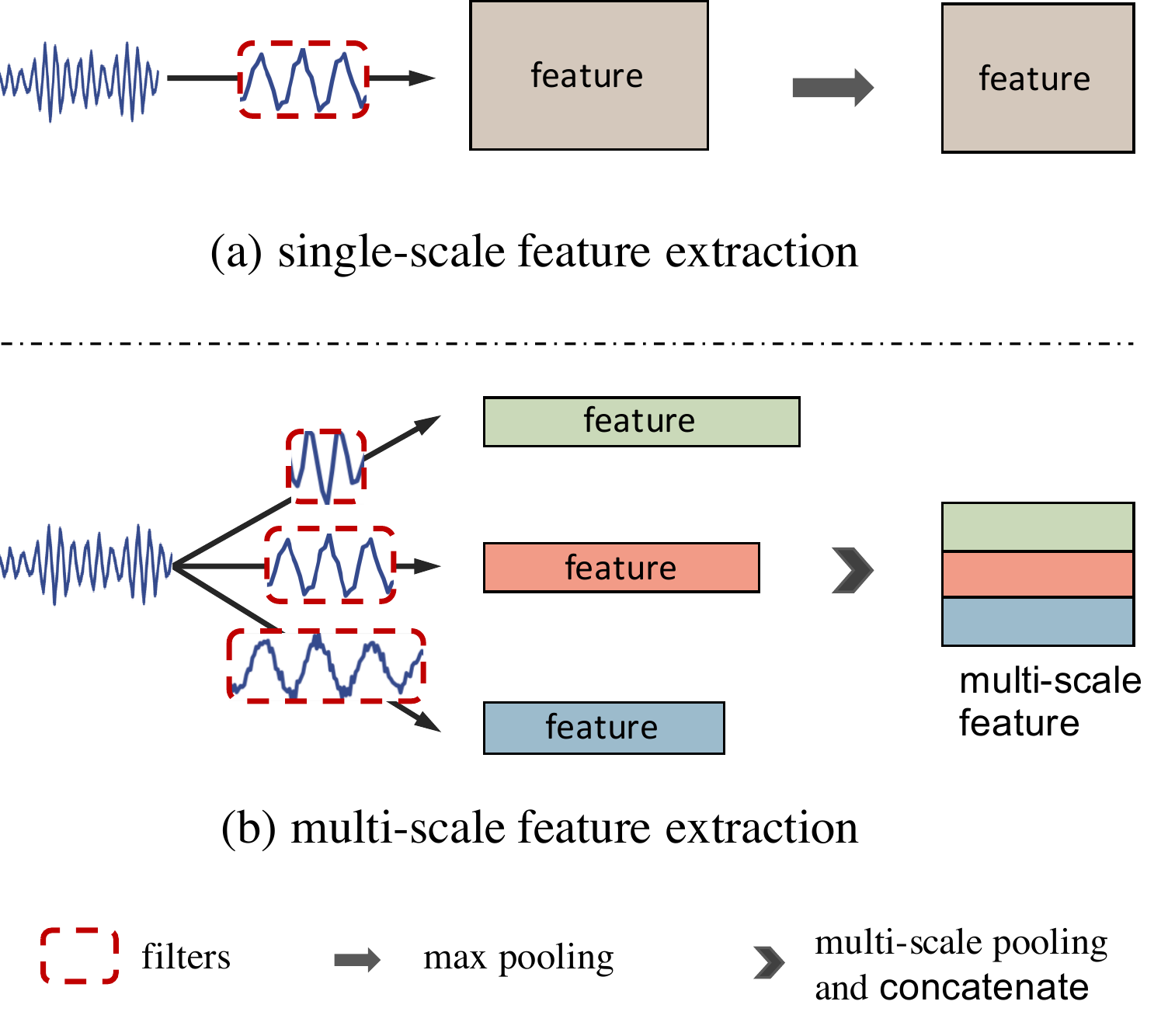}
\caption{single-scale vs. multi-scale feature extraction}
\label{fig:single_multi}
\end{figure}

\begin{figure*}[!t]
\setlength{\abovecaptionskip}{0pt}
\setlength{\belowcaptionskip}{10pt}
\centering
\includegraphics[width=1.0\textwidth]{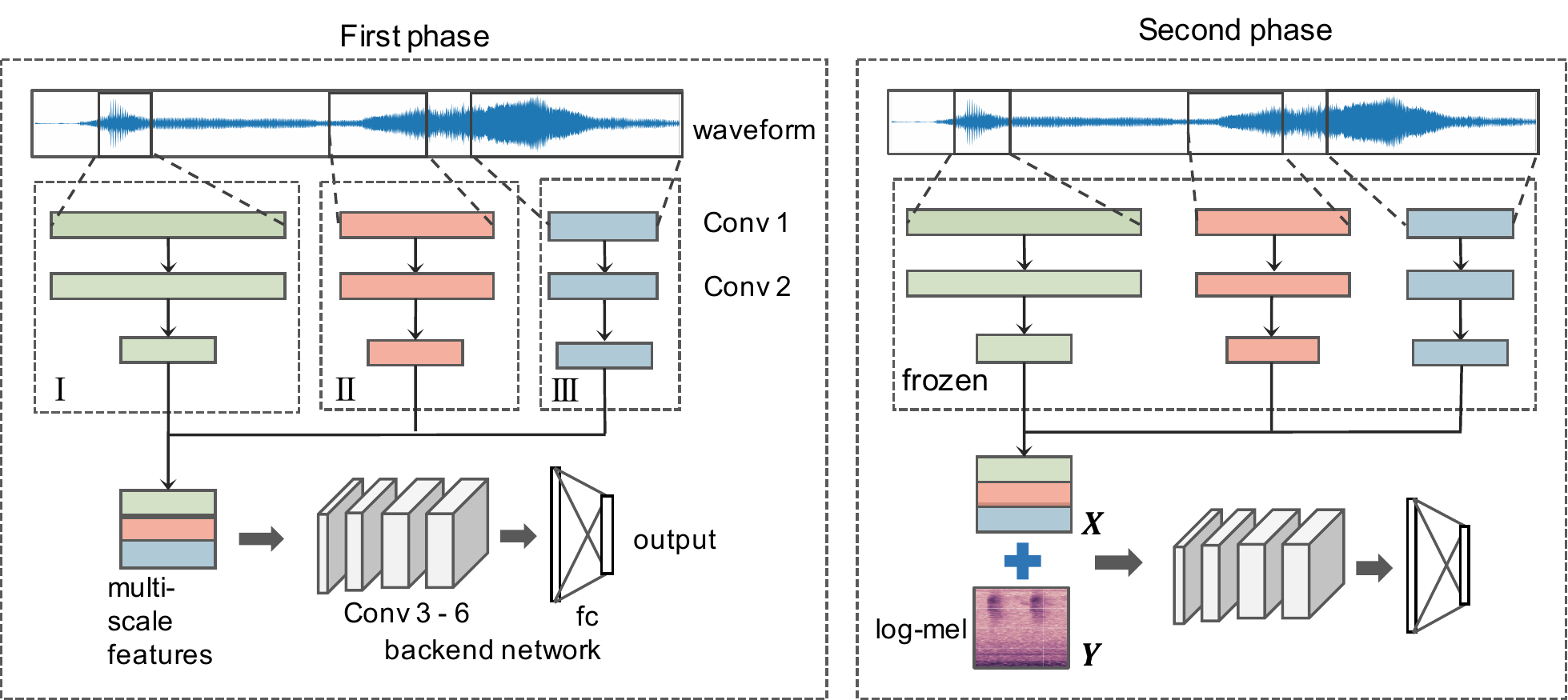}
\caption{\textbf{Network Architecture and Two-phases Training.} 
Architecture of WaveMsNet and two-phases training method for ESC task.}
\label{fig:model}
\end{figure*}

A popular approach which learns from waveform is passing the waveform through a time-domain convolution which has fixed filter size and followed by a pooling step to create invariance to phase shifts and further downsample the signal. This is the so-called single-scale model as Fig. \ref{fig:single_multi}(a) shows. However, features extracted by the single-scale model is not discriminative enough. 

First of all, although increasing the time and frequency resolution of the employed representation may be desirable, the uncertainty principle imposes a theoretical limit on how these two can be combined. When analyze a signal in time and frequency together, then more we zoom in into time, the equivalent amount we zoom out in frequency, and vice versa. On the other hand, frequency resolution is more critical for the classification task, if wide windows are employed on the waveform, we learn more about the low-frequency area, ignoring the high-frequency part, narrow windows behave in the opposite way. It is always a trade-off. Single-scale can not always balance them. 

Therefore, we propose the multi-scale convolution operation (Fig. \ref{fig:single_multi}(b)). At multi-scale convolution layer, the waveform signal $w(i)$ are convolved with filters $h^{(s)}$ at different scales $s$. We have that

\begin{equation}
x_j^{(s)}(n)=f(\sum_{i=0}^{N-1}w(i) h_j^{(s)}(n-i)+b_j^{(s)})
\label{equ:1}
\end{equation}
where $f$ is an activation function, $N$ is the length of $w(i)$ and $b_j$ is an additive bias. Three scales are chosen representatively ($s=1,2,3$). As thus, to learn high frequency features, filters with a short window are applied at a small stride on the waveforms. Low-frequency features, on the contrary, employ a long window that can be applied at a larger stride. At the same time, we wish to learn high frequency resolution through the long window (actually it does as shown in section \ref{sub:results_and_analysis}). Then feature maps at different scales are concatenate alone $frequency$ axis and a multi-scale max pooling is employed to downsize the feature map to the same dimension on the $time$ axis. 

% \begin{equation}
% \bm{X}=g(\bm{x}_1^{(1)},...,\bm{x}_j^{(1)},\bm{x}_1^{(2)},...,\bm{x}_j^{(2)},\bm{x}_1^{(3)},...,\bm{x}_j^{(3)})
% \label{equ:2}
% \end{equation}

\begin{table}[!b]
\centering
\caption{The layers configuration of WaveMsNet.}
\label{netconf}
\begin{tabular}{|c|c|c|c|c|}
\hline
layer name             & output size   & \begin{tabular}[c]{@{}c@{}}filter size,\\ filters number\end{tabular} & filter stride   & \begin{tabular}[c]{@{}c@{}}max\\ pooling\end{tabular}\\ 
\hline
Input     & 66150   & \multicolumn{3}{c|}{-} \\ 
\hline
Conv1     & \begin{tabular}[c]{@{}c@{}}\RNum{1}: 66150\\ \RNum{2}: 13230\\ \RNum{3}: 6615\end{tabular} & \begin{tabular}[c]{@{}c@{}}\RNum{1}: $11\times1$, 32\\ \RNum{2}: $51\times1$, 32 \\ \RNum{3}: $101\times1$, 32  \end{tabular}    & \begin{tabular}[c]{@{}c@{}}\RNum{1}: 1\\ \RNum{2}: 5\\ \RNum{3}: 10\end{tabular} & no pooling \\ 
\hline
\multirow{3}{*}{Conv2} & \multirow{3}{*}{441}  & \multirow{3}{*}{$11\times1$, 32}   & \multirow{3}{*}{1} & \multirow{3}{*}{\begin{tabular}[c]{@{}c@{}}\RNum{1}: 150\\ \RNum{2}: 30\\ \RNum{3}: 15\end{tabular}} \\ & & & & \\ & & & &\\ 
\hline
Conv3     & $32\times40$    & $3\times3$, 64    & $1\times1$  & $3\times11$ \\ 
\hline
Conv4     & $16\times20$    & $3\times3$, 128   & $1\times1$  & $2\times2$  \\ 
\hline
Conv5     & $8\times10$     & $3\times3$, 256   & $1\times1$  & $2\times2$  \\ 
\hline
Conv6     & $4\times5$      & $3\times3$, 256   & $1\times1$  & $2\times2$  \\ 
\hline
Fc        & 4096            & \multicolumn{3}{c|}{4096, dropout: 50\%}      \\ 
\hline

\end{tabular}
\end{table}
% subsection Single-scale vs. Multi-scale Model (end)

\subsection{Feature Fusion} % (fold)
\label{sub:feature_fusion}
While we considering both the waveform and analytic signal transformation, one of the mainstream approaches is training several independent models and calculating the average of output probabilities of each model. The improvement of the classification performance after this simple combination (as shown in section \ref{ssub:two_phase_feature_fusion}) indicates that our multi-scale features have the capacity to complement the log-mel features.

As our experiments reveal later (Table \ref{acc_phase2}), simple combination (averaging the probabilities) does not make full use of the information in different features and the performance is not optimal. We explore the learning method to extract the complementary information between the different features in a single end-to-end model.

We propose a two-phase method of feature fusion, which aims at joining log-mel features in. In the first phase, a feature extractor will be trained and the multi-scale feature map $\bm{X}\in\mathbb{R}^{H \times W}$ will be extracted directly on the time-series waveforms, where $H$ and $W$ is the $freqency$ and $time$ dimension respectively. In the second phase, the same-dimension log-mel features $\bm{Y}\in\mathbb{R}^{H \times W}$ are stack on the waveform features to form a two-channels feature map. It is convolved with learnable kernels and put through the activation function $f$ to form the output feature map $\bm{O}_j\in\mathbb{R}^{H \times W}$
\begin{equation}
\bm{O}_j=f(\sum_{i\in M}\bm{X}_i*\bm{k}_{ij} + \sum_{i\in L}\bm{Y}_{i}*\bm{k}_{ij}+b_j)
\label{equ:3}
\end{equation}
where $M$ and $L$ represent selections of the multi-scale feature map and log-mel feature map respectively, the $j$th output map is produced by kernel $\bm{k}_{ij}$ and each output map is given an additive bias b. Then we fine-tune the backend network while keeping the parameters in the feature extractor fixed during the back propagation.

To overcome the vanishing gradient problem, we designed a shallow backend network. Deeper networks can extract more abstract features, but with the deepening of the network layer, small gradients have little effect on the weights in front of the network, even if residual connection is used (as experiment in section \ref{ssub:two_phase_feature_fusion}). By reducing the number of backend network layers, the first few layers could have a good ability to extract discrepant features and as convergent as possible.

% subsection feature_fusion (end)

\subsection{Neural Network Architecture}
According to the above, we propose the multi-scale convolutional neural network with feature fusion as Fig. \ref{fig:model} shows. Firstly, we apply the multi-scale convolution operation on the input waveform. Three scales are chosen: \RNum{1}: (size 11, stride 1), \RNum{2}: (size 51, stride 5), \RNum{3}: (size 101, stride 10). Each scale has 32 filters in the first layer. Another convolutional layer is followed to create invariance to phase shifts with filter size 11 and stride 1. We aggressively reduce the temporal resolution to 441 with a max pooling layer to each scale feature map. We use non-overlapping (the pooling stride is same as pooling size) max-pooling. Then, we concatenate three feature map together and we get the multi-scale feature map as shown in Fig. \ref{fig:model}.

Then, the backend network is applied on the multi-scale feature map which can be seen as time-frequency representation. Four convolutional layers are followed. We use small receptive field $3\times3$ in $frequency \times time$ for all layers. Small filter size reduces the number of parameters in each layer and control the model sizes and computation cost. We apply non-overlapping max pooling after convolutional layers. Finally, we apply a fully connected layers with 4096 neurons and an output layer which has as many neurons as the number of classes.

We adopt auxiliary layers called batch normalization (BN) \cite{Ioffe2015BatchNormalization} after each convolutional layer (before pooling layer). BN alleviates the problem of exploding and vanishing gradients, a common problem in optimizing deep architectures. It normalizes the output in one batch of the previous layer so the gradients are well behaved. With BN layer applied, we can accelerate the training phase. 

% subsection Neural Network Architecture (end)

\subsection{Implementation Details} % (fold)
We use the exactly same model in two training phases. When training, we randomly select a 1.5 seconds waveform as input which were employed by Tokozume \cite{Tokozume2017Learning}, in testing phase, we use the probability-voting strategy. Rectified Linear Units (ReLUs) have been applied for each layer. We use momentum stochastic gradient descent (momentum SGD) optimizer to train the network where momentum set as 0.9. We run each model for 180 epochs until convergence. Learning rate is set as $10^{-2}$ for first 50 epochs, $10^{-3}$ for next 50 epochs, $10^{-4}$ for next 50 epochs and $10^{-5}$ for last 30 epochs. The weights in each model are initialized from scratch without any pre-trained model or Gammatone initialization because we want to learn the complements feature with handcraft features such as mel-scale feature. Dropout layers are followed full connected layer with a dropout rate of 0.5 to avoid overfitting. All weight parameters are subjected to $\ell_2$ regularization with coefficient $5\times10^{-4}$.
% subsection implementation_details (end)

% subsection multiscale_learning_on_waveform_processing (end)

\section{Experiments}

\subsection{Datasets} % (fold)
\label{sub:Dataset}
ESC-50 and ESC-10 datasets which are public labeled sets of environmental recordings are used in our experiments. ESC-50 dataset comprises 50 equally balanced classes, each clip is about 5 seconds and sampled at 44.1kHz. The 50 classes can be divided into 5 major groups: animals, natural soundscapes and water sounds, human non-speech sound, interior/domestic sounds, and exterior/urban noises. The dataset provides an exposure to a variety of sound sources, some very common (\emph{laughter, cat meowing, dog barking}), some quite distinct (\emph{glass breaking, brushing teeth}) and then some where the differences are more nuanced (\emph{helicopter and airplane noise}). The ESC-10 is a selection of 10 classes from the ESC-50 dataset. Datasets have been prearranged into 5 folds for comparable cross-validation and other experiments \cite{Piczak2015Environmental,Tokozume2017Learning, Aytar2016SoundNet} used these folds. For a fair comparison, the same folds’ division are proposed in our evaluation. The audios are not down-sampled, because we want to keep more high frequency information. We shuffle the training data but do not perform any data augmentation. 
% subsection Dataset (end)

\subsection{Effectiveness of multi-scale convolution operation} % (fold)

\begin{table}[b]
\setlength{\abovecaptionskip}{0pt}
\setlength{\belowcaptionskip}{10pt}
\centering
\caption{Comparison of the Multi-scale and Single-scale Model}
\begin{threeparttable}
\begin{tabular}{lccccc}
\hline
\multirow{2}{*}{Model} & \multicolumn{3}{c}{Filter numbers}   &  \multicolumn{2}{c}{Mean accuracy(\%)}   \\
                       & Scale \RNum{1} & Scale \RNum{2} & Scale \RNum{3}  & ESC-10  & ESC-50   \\ 
\hline
SRF                    & 96    & 0     & 0      & 85.85   & 67.10 \\
MRF                    & 0     & 96    & 0      & 86.70   & 67.55 \\
LRF                    & 0     & 0     & 96     & 85.75   & 67.05 \\
\hline
WaveMsNet              & 32    & 32    & 32     & \textbf{88.05}   & \textbf{70.05} \\
\hline
\end{tabular}
\end{threeparttable}
\label{multiscale}
\end{table}

We hypothesize that applying multi-scale convolution operation could allow each scale to learn filters selective to the frequencies that it can most efficiently represent. To test our hypothesis, we train variant models in single scales. We compare the performance with constant filter size at three different scales, small receptive field model (SRF), middle receptive field model (MRF), and large receptive field model (LRF). These three models remain only one corresponding scale (SRF remains scale \RNum{1}, MRF remains scale \RNum{2} and LRF remains scale \RNum{3}) and use triple filters in Conv1 and Conv2 layers for fair comparison with multi-scale. These three variant models are trained separately. The input and the rest of network is same as Fig. \ref{fig:model}

Table \ref{multiscale} shows the accuracies using multi-scale features and single-scale features when we only take waveform as input. Experiments conduct with different filter sizes, strides and pooling size in the CNN training. We observe that our WaveMsNet substantially improve the performance (88.05\% and 70.05\%). We have an improvement of 2.95\%, 2.50\%, 3.00\% compared with SRF, MRF, and LRF model respectively with same filters number on ESC-50 dataset. Also, multi-scale model achieves at lest 1.35\% improvement on ESC-10. To our knowledge, it is the best-performing end-to-end model which only use waveform as input. The significantly improvement from single-scale to multi-scale proves that different scales have learned more discriminative features from raw waveform. Further, the improvement is more notable on the larger dataset (ESC-50), which implies that our model has good generalization ability. 

\begin{figure}[tb]
\setlength{\abovecaptionskip}{0pt}
\setlength{\belowcaptionskip}{10pt}
\centering
\includegraphics[scale=.7]{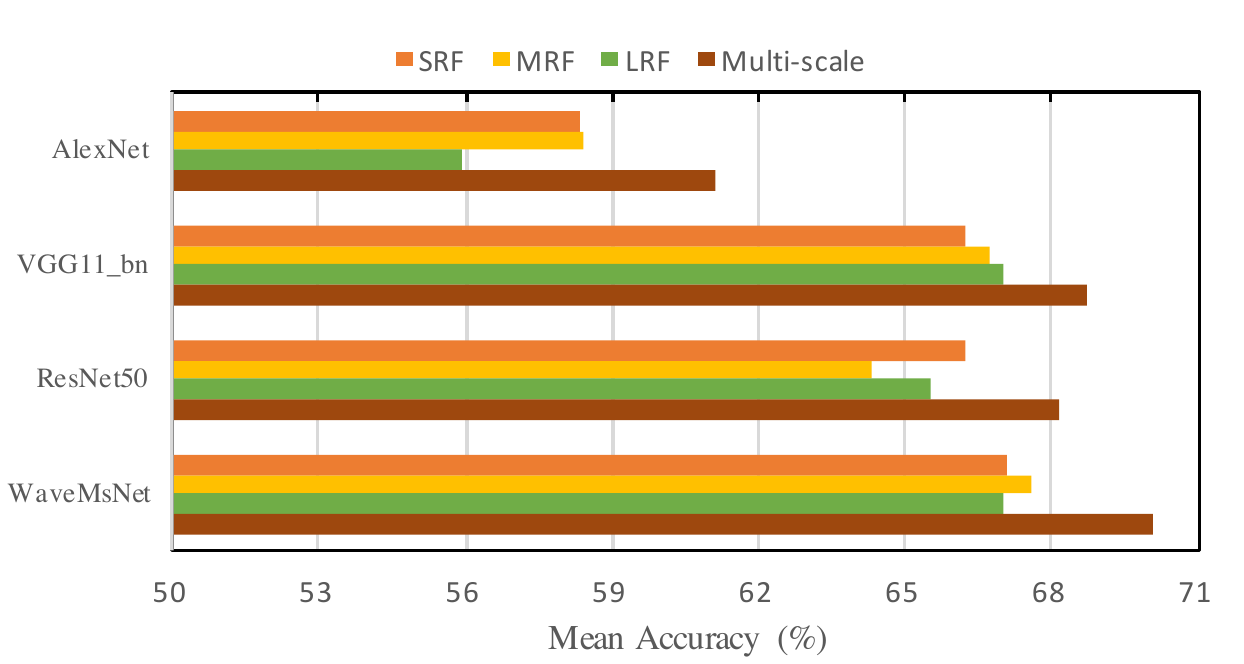}
\caption{\textbf{Effectiveness of multi-scale models.}
To further verify the effectiveness of the multi-scale. We compare WaveMsNet and other three backend networks: AlexNet, VGG, ResNet. On ESC-50, Multi-scale models are superior to single-scale ones regardless of the backend networks.}
\label{fig:otherNet}
\end{figure}

To further verify the effectiveness of the multi-scale convolution operation, we employ different backend networks, all of which are widely used and well-preformed in the field of image. They are AlexNet \cite{AlexNet}, VGG (11 layers with BN) \cite{VGG} and ResNet (50-layers) \cite{ResNet}. As Fig. \ref{fig:otherNet} demonstrates, the multi-scale models consistently outperform single-scale models. It indicates that multi-scale models have a wide range of effectiveness. In our shallow network, vanishing gradient problem is well suppressed during back-propagating. The first two layers of our network can converge better so that our WaveMsNet matches or even exceeds other very deep convolutional neural network.
% subsection Effectiveness of multi-scale features (end)

\begin{figure*}[!t]
\setlength{\abovecaptionskip}{0pt}
\setlength{\belowcaptionskip}{10pt}
\centering
\includegraphics[scale=.5]{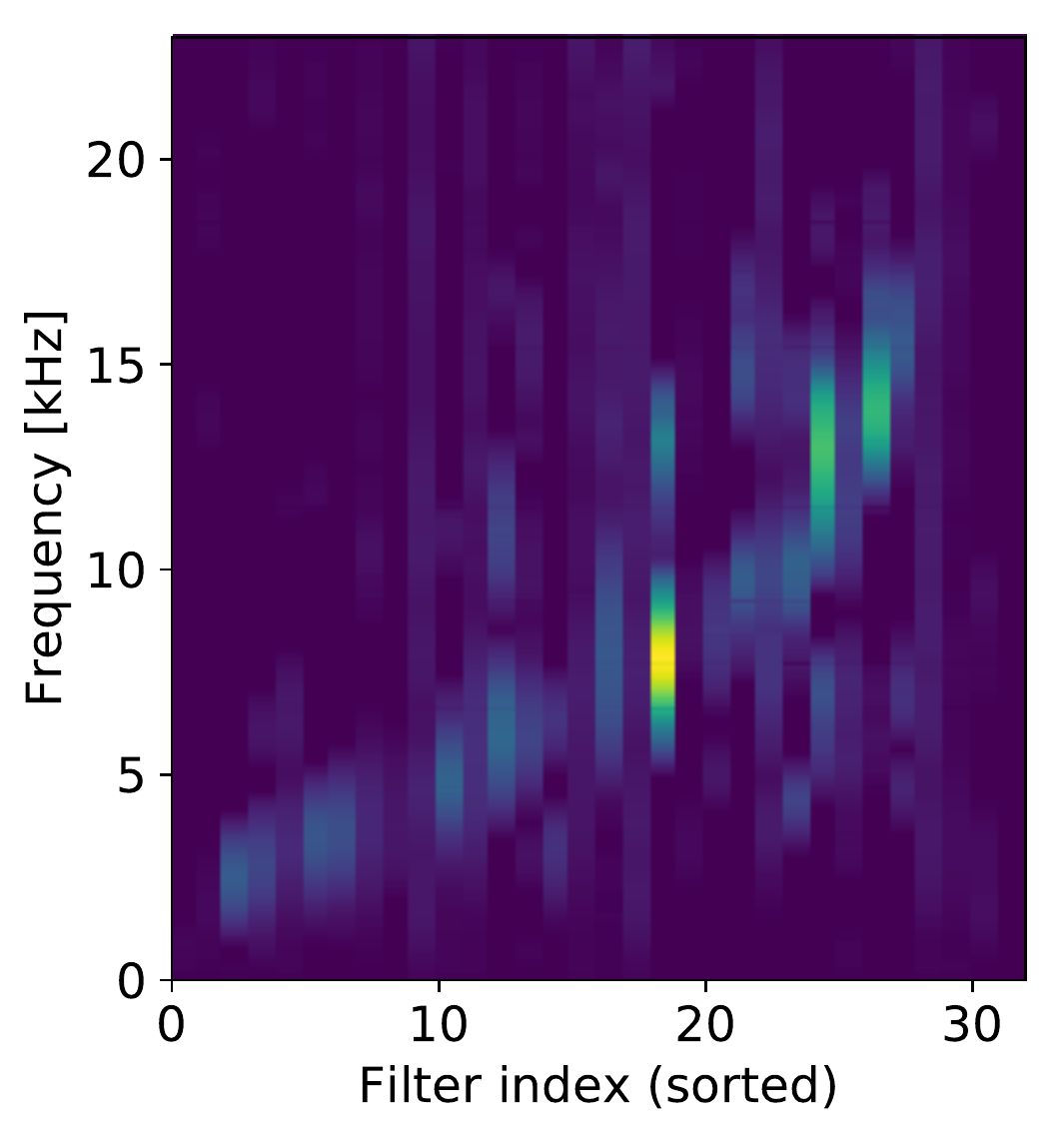}
\includegraphics[scale=.5]{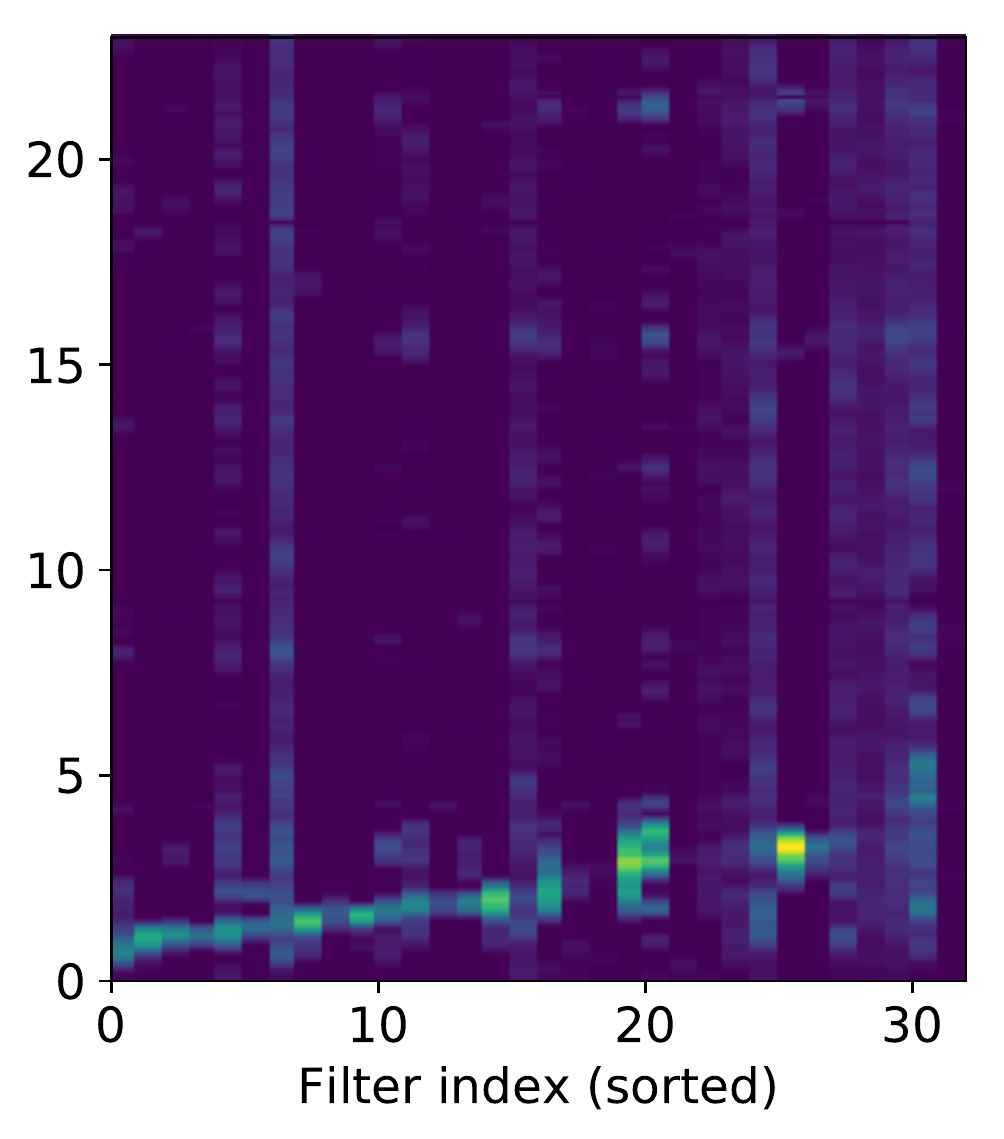}
\includegraphics[scale=.5]{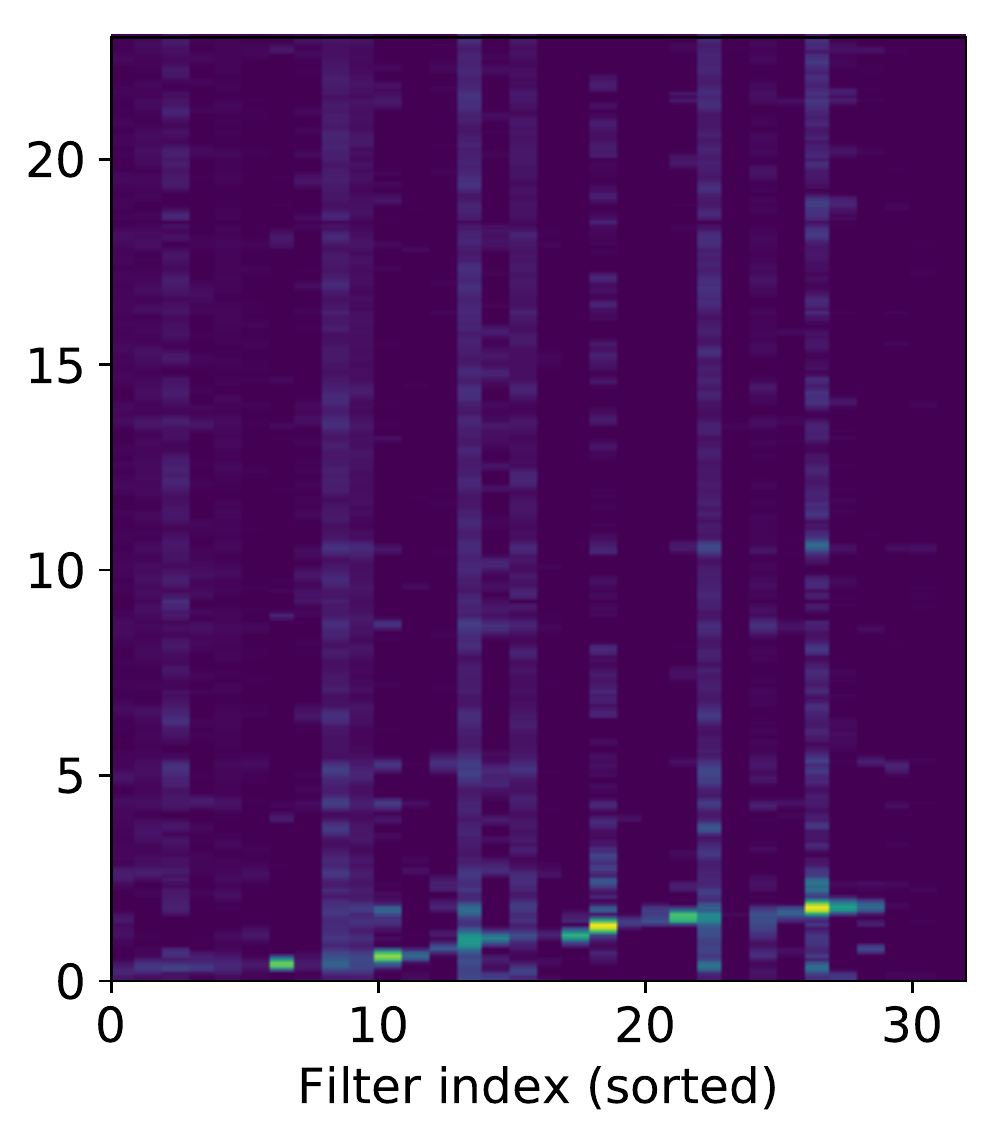}
\caption{\textbf{Frequency response of the multi-scale feature maps. } 
Left shows the frequency response of feature map product by scale \RNum{1}. Middle corresponds to scale \RNum{2}. Right corresponds to scale \RNum{3}. }
\label{fig:response}
\end{figure*}

\subsection{Two-phase feature fusion} % (fold)
\label{ssub:two_phase_feature_fusion}

\begin{table}[b]
\centering
\begin{threeparttable}
\caption{Performances with different training methods}

\begin{tabular}{lcc}
\hline
\multirow{2}{*}{Method}   & \multicolumn{2}{c}{Accuracy (\%)$\pm$std} \\
                          & ESC-10          & ESC-50                  \\
\hline
after first phase               & 88.05 $\pm$ 2.50  & 70.05 $\pm$ 0.63          \\
simple combination(two models)  & 89.50 $\pm$ 1.23  & 78.50 $\pm$ 1.50          \\
after second phase (not frozen) & 92.50 $\pm$ 1.87  & 77.85 $\pm$ 2.10          \\
after second phase(frozen)      & \textbf{93.75} $\pm$ 0.63  & \textbf{79.10} $\pm$ 1.63          \\
one-phase training method       & 88.75 $\pm$ 2.12  & 76.35 $\pm$ 1.25          \\
\hline
\end{tabular}
\label{acc_phase2}
\end{threeparttable}
\end{table}

We train our backend network with the log-mel features to form a new network and simply combine it with WaveMsNet we got above -- calculating the average of output probabilities of each model. As shown in Table \ref{acc_phase2}, we get 89.50\% and 78.50\% accuracy on ESC-10 and ESC-50. The improvement on accuracy indicates that the multi-scale features we have extracted from waveforms are highly complementary from log-mel features.

Now, we use the two-phase method to fuse the features. In first phase, we only use waveform as input to train WaveMsNet. In second phase, we fuse log-mel features with the multi-scale features as we introduced in section \ref{sub:feature_fusion}. We fine-tune the layers after Conv3 and keep the front part of the network (Conv1 and Conv2) freezing. That is to say, in the second training phase, parameters will only update in layers after Conv3 and not update in Conv1 and Conv2. 

Table \ref{acc_phase2} shows results after the second phase and compares the performance when the Conv1 and Conv2 are frozen and not frozen. The accuracies have been greatly improved 5.70\% and 9.05\% on ESC-10 and ESC-50 datasets after second phase. In comparison, the improvement is 4.45\% and 7.85\% on these two datasets when we keep the parameters in Conv1 and Conv2 updatable. The improvements are more obvious at second phase if we freeze the parameters than not. We infer that it’s because log-mel features will disturb the network’s front part which aiming at extracting features from waveform. Furthermore, we compare with one-phase training method which fuse the multi-scale features with log-mel features from the beginning. The performances degrade 5.00\% and 2.75\% respectively because the layers would be optimized under different hyper-parameters. The network would hard to take into account the waveform and log-mel if we unify the two training phase into one. The two-phase method combines waveform-based features with spectral-based features in a single model instead of training two separate models as previous works did and outperforms others.
% subsection Two-phase feature fusion (end)

\subsection{Results and Analysis} % (fold)
\label{sub:results_and_analysis}

\begin{table}[tb]
\centering
\begin{threeparttable}
\caption{Comparison of Accuracy of ESC Dataset}
\begin{tabular}{lcc}
\hline
\multirow{2}{*}{Model}   & \multicolumn{2}{c}{Mean accuracy(\%)} \\
                          & ESC-10          & ESC-50                  \\
\hline
Piczak’s CNN\cite{Piczak2015Environmental}                     & 81.00    & 64.50        \\
Tokozume’s Logmel-CNN\cite{Tokozume2017Learning}               &   -      & 66.50        \\
EnvNet\cite{Tokozume2017Learning}                              &   -      & 64.00        \\
D-CNN-ESC\cite{Zhang2017Dilated}                               &   -      & 68.10        \\
AlexNet\cite{Boddapati2017Classifying}                         &  85.00   & 69.00        \\
GoogLeNet\cite{Boddapati2017Classifying}                       &  90.00   & 73.20        \\
Aytar\cite{Aytar2016SoundNet}                                  &  92.20   & 74.20        \\
EnvNet $\oplus$ Logmel-CNN\cite{Tokozume2017Learning}\tnote{*} &    -     & 71.00        \\
WaveMsNet after first phase(ours)  & \textbf{88.05}  & \textbf{70.05}        \\
WaveMsNet after second phase(ours) & \textbf{93.75}  & \textbf{79.10}        \\
\hline
Human performance \cite{ESC_Dataset}                           & 95.70    & 81.30        \\
\hline
\end{tabular}
\begin{tablenotes}\small 
\item[*] The $\oplus$ sign indicates system simple combination before softmax
\end{tablenotes}
\label{results_on_ESC50}
\end{threeparttable}
\end{table}

Our proposed work is compared with the other studies in literature in Table \ref{results_on_ESC50}. WaveMsNet trained after two phase get an accuracy of 93.75\% and 79.10\% on ESC-10 and ESC-50 which about to match the performance of untrained human participants on this datasets (95.70\% and 81.3\%).

Fig. \ref{fig:response} shows the responses of the multi-scale feature maps. Most of the filters learn to be band-pass filters. Scale \RNum{1} has learned more dispersed bands across the frequency that can extract feature from all frequency and the trend of the center frequency matches the mel-scale (how human perceive the sound). But the frequency resolution is lower. On the contrary, scale \RNum{3} has learned high frequency resolution bands and most of them locate at the low frequency area. But it does not have sufficient filters in the high frequency range. Scale \RNum{2} behaves between scale \RNum{1} and \RNum{3}. This indicts that different scales could learn discrepant features and the filter banks split responsibilities based on what they efficiently can represent. This explains why multi-scale models get a better performance than single-scale models shown in Table \ref{multiscale}.
% subsection results_and_analysis (end)

\section{Conclusion}
In this paper, we proposed a multi-scale CNN (WaveMsNet) that operates directly on raw waveform inputs. We presented our model to learn more efficient representations by multiple scales. It achieves the state-of-the-art model using only raw waveform as input. By the two-phase training method, we fused the waveform and log-mel features in a single model and got a significant improvement in classification accuracy on ESC-10 and ESC-50 datasets. Furthermore, we analyzed the discrimination of features learning at different scales and had insight into the reasons of performance improvement.

\section*{Acknowledgment}
This work is supported by the National Key Research and Development Program of China (2016YFB1000100).

\bibliographystyle{IEEEtran}
\bibliography{bib_ref}

\begin{thebibliography}{10}
\providecommand{\url}[1]{#1}
\csname url@samestyle\endcsname
\providecommand{\newblock}{\relax}
\providecommand{\bibinfo}[2]{#2}
\providecommand{\BIBentrySTDinterwordspacing}{\spaceskip=0pt\relax}
\providecommand{\BIBentryALTinterwordstretchfactor}{4}
\providecommand{\BIBentryALTinterwordspacing}{\spaceskip=\fontdimen2\font plus
\BIBentryALTinterwordstretchfactor\fontdimen3\font minus
  \fontdimen4\font\relax}
\providecommand{\BIBforeignlanguage}[2]{{%
\expandafter\ifx\csname l@#1\endcsname\relax
\typeout{** WARNING: IEEEtran.bst: No hyphenation pattern has been}%
\typeout{** loaded for the language `#1'. Using the pattern for}%
\typeout{** the default language instead.}%
\else
\language=\csname l@#1\endcsname
\fi
#2}}
\providecommand{\BIBdecl}{\relax}
\BIBdecl

\bibitem{ESC_Dataset}
\BIBentryALTinterwordspacing
K.~J. Piczak, ``{ESC}: {Dataset} for {Environmental Sound Classification},'' in
  \emph{Proceedings of the 23rd {Annual ACM Conference} on {Multimedia}}.\hskip
  1em plus 0.5em minus 0.4em\relax {ACM Press}, pp. 1015--1018. [Online].
  Available: \url{http://dl.acm.org/citation.cfm?doid=2733373.2806390}
\BIBentrySTDinterwordspacing

\bibitem{Piczak2015Environmental}
------, ``Environmental sound classification with convolutional neural
  networks,'' in \emph{Machine Learning for Signal Processing (MLSP), 2015 IEEE
  25th International Workshop on}.\hskip 1em plus 0.5em minus 0.4em\relax IEEE,
  2015, pp. 1--6.

\bibitem{Dai2016Very}
W.~Dai, C.~Dai, S.~Qu, J.~Li, and S.~Das, ``Very deep convolutional neural
  networks for raw waveforms,'' in \emph{Acoustics, Speech and Signal
  Processing (ICASSP), 2017 IEEE International Conference on}.\hskip 1em plus
  0.5em minus 0.4em\relax IEEE, 2017, pp. 421--425.

\bibitem{Salamon2014A}
J.~Salamon, C.~Jacoby, and J.~P. Bello, ``A dataset and taxonomy for urban
  sound research,'' in \emph{ACM International Conference on Multimedia}, 2014,
  pp. 1041--1044.

\bibitem{Tokozume2017Learning}
Y.~Tokozume and T.~Harada, ``Learning environmental sounds with end-to-end
  convolutional neural network,'' in \emph{IEEE International Conference on
  Acoustics, Speech and Signal Processing}, 2017, pp. 2721--2725.

\bibitem{Boddapati2017Classifying}
V.~Boddapati, A.~Petef, J.~Rasmusson, and L.~Lundberg, ``Classifying
  environmental sounds using image recognition networks,'' \emph{Procedia
  Computer Science}, vol. 112, pp. 2048--2056, 2017.

\bibitem{Zhang2017Dilated}
X.~Zhang, Y.~Zou, and W.~Shi, ``Dilated convolution neural network with
  leakyrelu for environmental sound classification,'' in \emph{International
  Conference on Digital Signal Processing}, 2017, pp. 1--5.

\bibitem{randomForest}
A.~Liaw, M.~Wiener \emph{et~al.}, ``Classification and regression by
  randomforest,'' \emph{R news}, vol.~2, no.~3, pp. 18--22, 2002.

\bibitem{Tara2015cldnn}
T.~N. Sainath, O.~Vinyals, A.~Senior, and H.~Sak, ``Convolutional, long
  short-term memory, fully connected deep neural networks,'' in \emph{IEEE
  International Conference on Acoustics, Speech and Signal Processing}, 2015,
  pp. 4580--4584.

\bibitem{Tara2016vad}
R.~Zazo, T.~N. Sainath, G.~Simko, and C.~Parada, ``Feature learning with
  raw-waveform cldnns for voice activity detection,'' in \emph{INTERSPEECH},
  2016, pp. 3668--3672.

\bibitem{Tara2015front-end}
A.~S. K. W.~W. T.~N.~Sainath, R. J.~Weiss and O.~Vinyals, ``Learning the speech
  front-end with raw waveform cldnns,'' in \emph{Sixteenth Annual Conference of
  the Internatianal Speech Communication Association}, 2015.

\bibitem{Tradition1}
D.~Barchiesi, D.~Giannoulis, S.~Dan, and M.~D. Plumbley, ``Acoustic scene
  classification: Classifying environments from the sounds they produce,''
  \emph{IEEE Signal Processing Magazine}, vol.~32, no.~3, pp. 16--34, 2015.

\bibitem{Tradition2}
J.~T. David~Li and D.~Toub, ``Auditory scene classification using machine
  learning techniques,'' \emph{AASP Challenge}, 2013.

\bibitem{Tradition3}
A.~Rakotomamonjy and G.~Gasso, ``Histogram of gradients of time-frequency
  representations for audio scene classification,'' \emph{IEEE/ACM Transactions
  on Audio Speech \& Language Processing}, vol.~23, no.~1, pp. 142--153, 2015.

\bibitem{Tradition4}
S.~Dan, D.~Giannoulis, E.~Benetos, M.~Lagrange, and M.~D. Plumbley, ``Detection
  and classification of acoustic scenes and events,'' \emph{IEEE Transactions
  on Multimedia}, vol.~17, no.~10, pp. 1733--1746, 2015.

\bibitem{Tradition5}
G.~Roma, W.~Nogueira, and P.~Herrera, ``Recurrence quantification analysis
  features for environmental sound recognition,'' \emph{Annals of the Rheumatic
  Diseases}, vol.~51, no.~9, pp. 1056--62, 2013.

\bibitem{Tradition6}
J.~Salamon and J.~P. Bello, ``Unsupervised feature learning for urban sound
  classification,'' in \emph{IEEE International Conference on Acoustics, Speech
  and Signal Processing}, 2015, pp. 171--175.

\bibitem{SVMforspeaker}
W.~M. Campbell, D.~E. Sturim, and D.~A. Reynolds, ``Support vector machines
  using gmm supervectors for speaker verification,'' \emph{IEEE signal
  processing letters}, vol.~13, no.~5, pp. 308--311, 2006.

\bibitem{Salamon2016Deep}
J.~Salamon and J.~Bello, ``Deep convolutional neural networks and data
  augmentation for environmental sound classification,'' \emph{IEEE Signal
  Processing Letters}, vol.~PP, no.~99, pp. 1--1, 2016.

\bibitem{Vacher2014Sound}
M.~Vacher, J.-F. Serignat, and S.~Chaillol, ``Sound classification in a smart
  room environment: an approach using gmm and hmm methods,'' in \emph{The 4th
  IEEE Conference on Speech Technology and Human-Computer Dialogue (SpeD 2007),
  Publishing House of the Romanian Academy (Bucharest)}, vol.~1, 2007, pp.
  135--146.

\bibitem{Agrawal2017Novel}
D.~M. Agrawal, H.~B. Sailor, M.~H. Soni, and H.~A. Patil, ``Novel teo-based
  gammatone features for environmental sound classification,'' in
  \emph{European Signal Processing Conference}, 2017, pp. 1809--1813.

\bibitem{AlexNet}
A.~Krizhevsky, I.~Sutskever, and G.~E. Hinton, ``Imagenet classification with
  deep convolutional neural networks,'' in \emph{International Conference on
  Neural Information Processing Systems}, 2012, pp. 1097--1105.

\bibitem{VGG}
K.~Simonyan and A.~Zisserman, ``Very deep convolutional networks for
  large-scale image recognition,'' \emph{arXiv preprint arXiv:1409.1556}, 2014.

\bibitem{ResNet}
K.~He, X.~Zhang, S.~Ren, and J.~Sun, ``Deep residual learning for image
  recognition,'' in \emph{Computer Vision and Pattern Recognition}, 2016, pp.
  770--778.

\bibitem{Huang2016DenseNet}
G.~Huang, Z.~Liu, K.~Q. Weinberger, and L.~van~der Maaten, ``Densely connected
  convolutional networks,'' in \emph{Proceedings of the IEEE conference on
  computer vision and pattern recognition}, vol.~1, no.~2, 2017, p.~3.

\bibitem{Ronneberger2015UNet}
O.~Ronneberger, P.~Fischer, and T.~Brox, \emph{U-Net: Convolutional Networks
  for Biomedical Image Segmentation}.\hskip 1em plus 0.5em minus 0.4em\relax
  Springer International Publishing, 2015.

\bibitem{zhang2017airport}
P.~Zhang, X.~Niu, Y.~Dou, and F.~Xia, ``Airport detection on optical satellite
  images using deep convolutional neural networks,'' \emph{IEEE Geoscience and
  Remote Sensing Letters}, vol.~14, no.~8, pp. 1183--1187, 2017.

\bibitem{wang2018local}
Q.~Wang, Y.~Dou, X.~Liu, F.~Xia, Q.~Lv, and K.~Yang, ``Local kernel alignment
  based multi-view clustering using extreme learning machine,''
  \emph{Neurocomputing}, vol. 275, pp. 1099--1111, 2018.

\bibitem{Zhang2015Character}
X.~Zhang, J.~Zhao, and Y.~LeCun, ``Character-level convolutional networks for
  text classification,'' in \emph{Advances in neural information processing
  systems}, 2015, pp. 649--657.

\bibitem{Kim2015Character}
Y.~Kim, Y.~Jernite, D.~Sontag, and A.~M. Rush, ``Character-aware neural
  language models,'' \emph{Computer Science}, 2015.

\bibitem{Lee2017Sample}
J.~Lee, J.~Park, K.~L. Kim, and J.~Nam, ``Sample-level deep convolutional
  neural networks for music auto-tagging using raw waveforms,'' \emph{arXiv
  preprint arXiv:1703.01789}, 2017.

\bibitem{Aytar2016SoundNet}
Y.~Aytar, C.~Vondrick, and A.~Torralba, ``Soundnet: Learning sound
  representations from unlabeled video,'' in \emph{Advances in Neural
  Information Processing Systems}, 2016, pp. 892--900.

\bibitem{Ioffe2015BatchNormalization}
S.~Ioffe and C.~Szegedy, ``Batch normalization: Accelerating deep network
  training by reducing internal covariate shift,'' in \emph{International
  conference on machine learning}, 2015, pp. 448--456.

\end{thebibliography}

\IEEEpeerreviewmaketitle
\end{document}